\documentclass[%
 preprint,
superscriptaddress,
 amsmath,amssymb,
 aps,
prb,
onecolumn,
]{revtex4-2}

\usepackage[version=4]{mhchem}
\usepackage{xcolor}
\usepackage{graphicx}
\usepackage{dcolumn}
\usepackage{bm}
\usepackage{placeins}
\usepackage{appendix}
\usepackage[utf8]{inputenc}
\usepackage[english]{babel}
\usepackage{float}

\begin{document}

\preprint{APS/123-QED}

\title{An inelastic neutron scattering study of the magnetic field dependence of the quantum dipolar garnet: Yb$_3$Ga$_5$O$_{12}$.}
\author{Edward Riordan}
\affiliation{Institut N\'{e}el, CNRS \& Université Grenoble Alpes, 38000 Grenoble, France}

\author{Monica Ciomaga Hatnean}
\affiliation{
PSI Center for Neutron and Muon Sciences, Paul Scherrer Institute, Villigen PSI, Switzerland.
} 

\author{Geetha Balakrishnan}
\affiliation{
Department of Physics, University of Warwick, Coventry CV4 7AL, UK
}%
\author{Kim Lefmann}
\affiliation{Nanoscience Center, Niels Bohr Institute, University of Copenhagen, Universitetsparken 5, DK-2100 Copenhagen \O , Denmark}
\author{Jacques Ollivier}
\affiliation{Institut Laue Langevin, CS 20156, 38042 Grenoble, France}
\author{Stephane Raymond}
\affiliation{Université Grenoble Alpes, CEA, IRIG, MEM, MDN, 38000 Grenoble, France}
\author{Elsa Lhotel}
\affiliation{Institut N\'{e}el, CNRS \& Université Grenoble Alpes, 38000 Grenoble, France}
\author{Pascale P. Deen}
\affiliation{Nanoscience Center, Niels Bohr Institute, University of Copenhagen, Universitetsparken 5, DK-2100 Copenhagen \O , Denmark}
\affiliation{European Spallation Source ERIC, 22363 Lund, Sweden}
\affiliation{Corresponding author: pascale.deen@ess.eu}

\date{\today}

\begin{abstract}

The garnet compound Yb$_{3}$Ga$_{5}$O$_{12}$  is a fascinating material that is considered highly suitable for low-temperature refrigeration, via the magnetocaloric effect, in addition to enabling the exploration of quantum states with long-range dipolar interactions. It has previously been theorized that the magnetocaloric effect can be enhanced, in Yb$_{3}$Ga$_{5}$O$_{12}$ , via magnetic soft mode excitations which in the hyperkagome structure would be derived from an emergent magnetic structure formed from nanosized 10-spin loops.

We study the magnetic field dependence of bands of magnetic soft mode excitations in the effective spin S = 1/2 hyperkagome compound Yb$_{3}$Ga$_{5}$O$_{12}$  using single crystal inelastic neutron scattering. We probe the magnetically short ranged ordered state, in which we determine magnetic nanoscale structures coexisting with a fluctuating state, and the magnetically saturated state, from which we determine the relevant magnetic interactions. We determine that Yb$_{3}$Ga$_{5}$O$_{12}$  can be described as a quantum dipolar magnet with perturbative weak near- neighbor and inter-hyperkagome exchange interaction. The magnetic excitations, under the application of a magnetic field, reveal highly robust soft modes with distinctive signatures of the quantum nature of the Yb$^{3+}$ spins. Our results enhance our understanding of soft modes in topological frustrated magnets that drive both the unusual physics of quantum dipolar systems and future refrigerant material design.

\end{abstract}

\maketitle


\section{Introduction} \label{sec:intro}

The interplay between geometry, topology, and electronic and magnetic states generates rich and complex many-body physics driven by topological flat bands with highly degenerate states. Topological flat band systems, in particular the electronic bands in crystalline solids, have been proposed to drive the physics of strongly correlated electronic phenomena as found in quantum spin liquid states \cite{TopologicalFlatBands_QuantumSpinLiquid}, unconventional superconductivity \cite{TopologicalFlatBands_SC}, and lattice versions of the fractional quantum Hall states \cite{TopologicalFlatBands_HallEffect}. 

Bands of dispersionless spin waves, also known as flat bands or soft modes, are hallmark features in some models of topological frustrated magnets that can drive exotic states of matter \cite{Chernyshev_2015}. Soft modes have been uncovered in 2 dimensional (2D) kagome and 3 dimensional (3D) hyperkagome lattices. In the low temperature phase of iron jarosite, ${\mathrm{KFe}}_{3}(\mathrm{OH}{)}_{6}({\mathrm{SO}}_{4}{)}_{2}$, the underlaying 2D kagome lattice, of triangles and hexagon loops, enable a large ground state degeneracy leading to flat bands with strong quantum effects even though the system is classical  \cite{Matan_2006,Chernyshev2015}. In the case of the 3D hyperkagome lattice, as observed in Gadolinium Gallium Garnet, Gd$_{3}$Ga$_{5}$O$_{12}$ (GGG), and possibly in Ytterbium Gallium Garnet, Yb$_{3}$Ga$_{5}$O$_{12}$ (YbGG), flat bands are derived from trimer, ten-ion loops and chain spin states that develop cooperative magnetic multipolar degrees of freedom from the underlying spin correlations \cite{Paddison_2015_GGG,Ambrumenil2015, YbGG_Lise_2021_PhysRevB.104.064425}. Flat magnetic modes can strongly affect the thermodynamic properties of materials. Indeed it is theoretically estimated that the rate of adiabatic temperature change in magnetocaloric materials can be improved by an order of magnitude in materials  with a high degree of soft modes \cite{Zhitomirksy2003,Dutton_2023}. As such, both GGG and YbGG are considered excellent materials for cryogenic magnetocaloric refrigeration \cite{PAIXAO_YbGG_2020}. 
 
 Magnetic flat band modes in GGG and YbGG reveal, upon early determination, a similar nature of the correlated state. Yet the exchange interactions are vastly different with GGG driven by nearest neighbor exchange and YbGG a possible quantum dipolar magnet reminiscent of the fascinating physics of spin-ice materials in which long range dipolar interactions drive exotic behaviors \cite{denHertog_PhysRevLett.84.3430, Fennell2009,  Edberg_PhysRevB.102.184408}. Beyond spin-ice physics, dipolar physics remains of great interest with dimensional effects driving a dipolar system from a ferromagnet to a complex spin textured state of interest for spintronics devices \cite{Politi_2006, Su_Dipolar2023}. 
 
 In this work we extend our study on the low temperature short ranged order phase and magnetically saturated phase of YbGG to understand and determine the relevant interactions that drive the emergent looped structures of which the flat modes can be the resultant signatures. We employ inelastic neutron scattering to study the magnetic field dependence of the magnetic excitation spectrum and uncover a robust state under applied magnetic field.

\section{Previous experimental magnetic signatures of YbGG.}

\begin{figure}[!htp]
    	\centering
		\includegraphics[width = 0.5\linewidth]{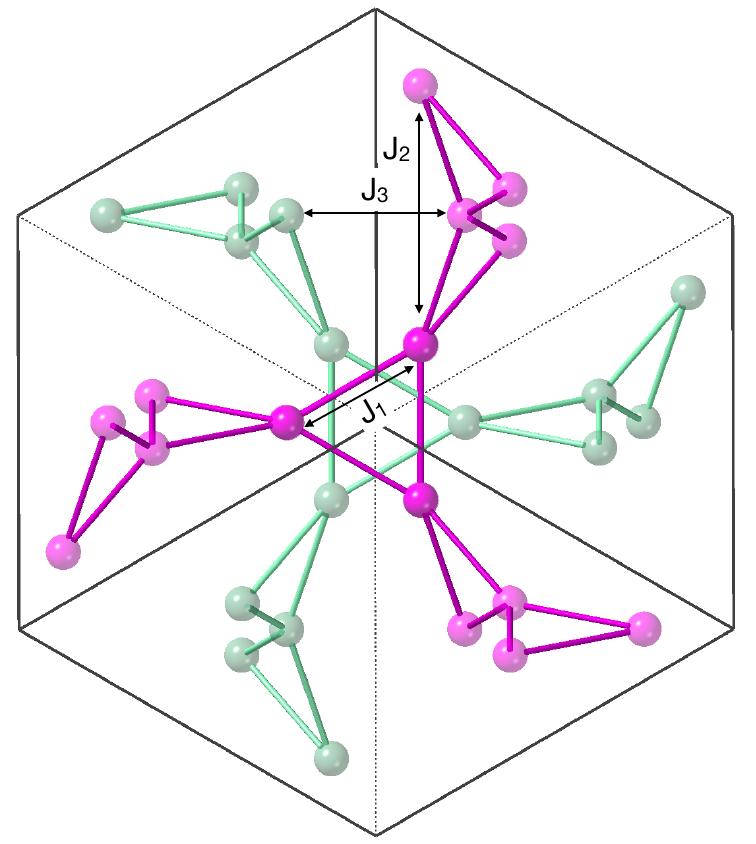}
        \caption{Position of Yb$^{3+}$ ions on the hyperkagome lattice viewed along the [1 1 1] crystallographic direction. $J_{1}$, $J_{2}$ represent the intra near-neighbor and next-nearest neighbor exchange in the hyperkagome lattice and $J_{3}$ is the interhyperkagome exchange.}
        \label{fig:Structure}
\end{figure}

In the 3D hyperkagome  crystal structure (Ia$\bar{3}$d) of YbGG, the Yb$^{3+}$ ions  are positioned on two interpenetrating triangular lattices with exchange interactions $J_{1}$ and $J_{2}$  representing the intra near-neighbor and  next-nearest neighbor exchange in a hyperkagome lattice and $J_{3}$ is the interhyperkagome exchange, see Fig. \ref{fig:Structure}. 

A large degree of magnetic frustration frequently develops when rare earth ions are arranged on the hyperkagome lattice, primarily due to the concomitant effect of antiferromagnetic near-neighbor exchange and anisotropy derived from dipolar interactions, single ion anisotropy and crystal field levels surrounding the rare earth site. In the case of YbGG, Yb$^{3+}$ ions are situated in a local orthorhombic crystal field environment splitting the $J=7/2$ magnetic state into four Kramers doublets with a ground state separated from the first excited level by 67 meV \cite{Pearson_1967}. The low temperature magnetic behaviour, $T<$ 5 K, is therefore entirely dominated by the ground state doublet and an effective quantum $S=1/2$ moment. Magnetically, YbGG reveals a very modest Curie-Weiss temperature. Yet slightly different values have been reported from ferromagnetic exchange with $\theta_{CW}$= 45 mK by Filippi {\it et al.} and $\theta_{CW}$ = 97 mK as reported by Lhotel {\it et al.} to antiferromagnetic exchange with $\theta_{CW}$= -21 mK as reported by Sandberg {\it et al.} \cite{Filippi_1980, lHotel2021PhysRevB.104.024427, YbGG_Lise_2021_PhysRevB.104.064425}. 

The local crystal field environment is slightly anisotropic with g-values $g_x$ = 2.84, $g_y$ = 3.59 and $g_z$ = -3.72 \cite{Carson1960, Pearson_1967}. However, for temperatures up to 4 K there is no anisotropy observed in magnetization measurements \cite{Filippi_1980}. Quantized magnetization plateaus, as a function of applied magnetic field, have been theoretically estimated and experimentally observed for a range of frustrated quantum spin lattices, inclusive of the kagome lattice with a spin $S=1/2$  \cite{Schulenburg_2002, Narumi2024}. However no such magnetization jumps have been observed for YbGG at the lowest temperatures, $T$ = 90 mK \cite{Filippi_1980}. Specific heat measurements \cite{Filippi_1980} reveal short range order developing below 600 mK with a broad peak centered around 200 mK, and a sharp lambda-shaped peak at $T_{n}$ = 54 mK, attributed initially to long range order \cite{Filippi_1980}. M\"{o}ssbauer spectroscopy could not identify any static magnetization below the lambda peak \cite{Hodges_2003}. However, neutron diffraction measurements were able to identify a magnetically ordered state with a propagation vector, \textbf{k} = ($\frac{1}{2}$ $\frac{1}{2}$ 0), albeit with a reduced value of the ordered moment \cite{Raymond2024}. The broad feature in $C_{v}$ accounts for 80\% of the total entropy while the lambda peak accounts for only $\sim$ 10\% \cite{Filippi_1980}. With increasing magnetic field, the broad peak in $C_{v}$(T) shifts to higher temperatures. This can be understood between 0.5 and 4 T with a fit to a Schottky anomaly but beyond that such fits are less well matched indicating more complex physics \cite{lHotel2021PhysRevB.104.024427}.

Magnetic fluctuations, measured by M\"{o}ssbauer spectroscopy,  linearly decrease down to 100 mK below which the magnetic fluctuations tend to a quasi-saturated value of 3 $\times$ 10$^{9}$ s$^{-1}$. The decrease in the spin relaxation rate followed by a saturation is similar to the behavior observed (by $\mu$SR) in GGG. These slow fluctuations indicate collective behavior of relatively large spin structures \cite{Jacobsen_2021}. 

Unusual soft mode magnetic excitations are observed in YbGG, at ambient magnetic fields at the lowest temperatures T $>$ T$_{n}$  with three low lying magnetic flat bands at $\Delta$E = 0.06(1) meV, 0.1 meV and 0.7 meV, termed INS1, INS2 and INS3 respectively  \cite{YbGG_Lise_2021_PhysRevB.104.064425,lHotel2021PhysRevB.104.024427}. Lhotel \textit{et} al. computed the magnon modes in YbGG under applied magnetic field and indicated that the excitation spectrum is derived from 12 excitation modes originating from the primitive unit cell that contains 12 Yb$^{3+}$ ions \cite{lHotel2021PhysRevB.104.024427}. These modes are gathered into several groups which cannot fully be resolved within the energy resolution of the instrument. 

In a magnetic field, the lowest energy branch of YbGG follows a Zeeman shift in accordance with field-dependent specific heat data \cite{lHotel2021PhysRevB.104.024427}. Three low lying flat band modes, with energies rather similar to those found in YbGG, also exist in GGG, however in GGG these modes do not follow a simple Zeeman magnetic field dependence. Theoretically it was determined, for GGG, that the dominant contribution to inelastic scattering at large momentum transfers comes from a band of almost dispersionless excitations corresponding to spin waves localized on a ten ion loop \cite{Ambrumenil2015}.

Lhotel \textit{et} al. \cite{lHotel2021PhysRevB.104.024427} were able to theoretically reproduce their experimentally obtained Curie-Weiss temperature of 97 mK using an isotropic $g$-value $\sim$ 3.42 and only dipole-dipole exchange to obtain a theoretical $\theta_{CW}$ = 101 mK. This certainly provides a strong indication that YbGG is a quantum dipolar magnet. In contrast, Sandberg \textit{et} al. reported that the specific heat signatures and elastic neutron scattering profiles could be closely reproduced by considering $J_{1}$, $J_{2}$, and dipolar exchange, $D$,  a $J_{1}J_{2}D$ model \cite{YbGG_Lise_2021_PhysRevB.104.064425}. The $J_{1}J_{2}D$ model reproduces many of the short range order elastic neutron scattering features as well as the specific heat measurements, in particular the short range correlations that develop below 600 mK and a lambda peak at 54 mK. However, some inconsistencies remain. For instance, certain aspects of the broad experimental neutron scattering dataset can be modelled using a $J_{1}$D model with ferromagnetic near-neighbor spin-spin correlations while other aspects of the data can be accurately reproduced using a $J_{1}J_{2}D$ model with antiferromagnetic near-neighbor spin-spin correlation. Extracting the spin structure across the 10-ion loop provides a director state, reminiscent of the director state found in GGG, with an easy axis along the local z-direction. These results are intriguing since Gd$^{3+}$ in GGG is a classical spin, with $S$=7/2, in which the dominant interactions are antiferromagnetic near neighbor exchange with a weak dipolar perturbation \cite{Paddison_2015_GGG} while Yb$^{3+}$ in YbGG, at the low temperature we are considering, can be considered a quantum spin.

Despite some significant insight into the ground state of YbGG the exact details have not yet been clarified and in particular the Hamiltonian for YbGG, with the main terms: 
\begin{equation}\label{JJDHamiltonian}
    \begin{split}
		\mathcal{H}=&J_1\sum_{\langle i,j\rangle} \, \mathbf{S}_i\cdot\mathbf{S}_j+J_2\sum_{\langle\langle i,j\rangle\rangle} \textbf{S}_i\cdot\textbf{S}_j
        +J_3\sum_{\langle\langle i,j\rangle\rangle} \textbf{S}_i\cdot\textbf{S}_j+\\
		&Da^3\sum_{i<j}\left(\frac{\mathbf{S}_i\cdot\mathbf{S}_j}{|\mathbf{r}_{ij}|^3}-3\frac{\left(\mathbf{S}_i\cdot\mathbf{r}_{ij}\right)\left(\mathbf{S}_j\cdot\mathbf{r}_{ij}\right)}{|\mathbf{r}_{ij}|^5}  \right), 
	\end{split}
\end{equation}
has not been fully determined. In equation (\ref{JJDHamiltonian}) $a$ is the nearest neighbor distance, $\textbf{r}_i$ is the position of the spin $\textbf{S}_i$ and $\textbf{r}_{ij}=\textbf{r}_i-\textbf{r}_j$. 

We thus employ single crystal inelastic neutron scattering to extract the magnetic field dependence of the double partial differential cross section $S(\textbf{Q}, \omega)$ accessing directly the spin-spin correlation function, associated dynamical information and to determine the relevant exchange parameters \cite{Boothroyd}. 



 
\section{Experimental Details}

Inelastic neutron scattering spectra have been measured as a function of applied magnetic field using IN5, the cold chopper spectrometer of the Institut Laue Langevin (ILL), on a high quality single crystal of YbGG previously studied \cite{YbGG_Lise_2021_PhysRevB.104.064425, IN5Data_2021}. 

The YbGG single crystal was aligned with the scattering plane in the  [$\zeta$ $\zeta$ -2$\zeta$] and [-$\xi$ $\xi$ 0] crystalline direction and cooled to 40 mK, as measured close to the sample. The long range order detected in specific heat measurements at 54 mK is not observed in these measurements which we attribute to incomplete thermalisation, since YbGG is a highly insulating material. The short range magnetic scattering observed indicates that the sample temperature is close to 200 mK for which a broad maximum in $C_{v}$ is observed. As such, the low temperature measurements are denoted by $T_{SRO}$. 
Magnetic fields, $\mu_{0}H$ = 0, 1, 2 and 5 T, have been applied along the [1 1 1] crystalline direction for an incident neutron energy $E_{i}$ = 3.27 meV with an instrumental energy resolution, $\delta$E, of 0.10 meV at energy transfer $\Delta E$ = 0.0 meV and $\delta$E = 0.08 meV at $\Delta E$ = 1.0 meV. Higher energy resolution measurements were performed with an incident neutron energy $E_{i}$ = 1.67 meV, $\delta E$ = 0.03 meV, for an applied magnetic field of $\mu_{0} H$ = 0 T and 0.5 T.

\section{Results}

\subsection{Overview}
\begin{figure}[!htp]
    	\centering
    \includegraphics[width=0.8\linewidth]{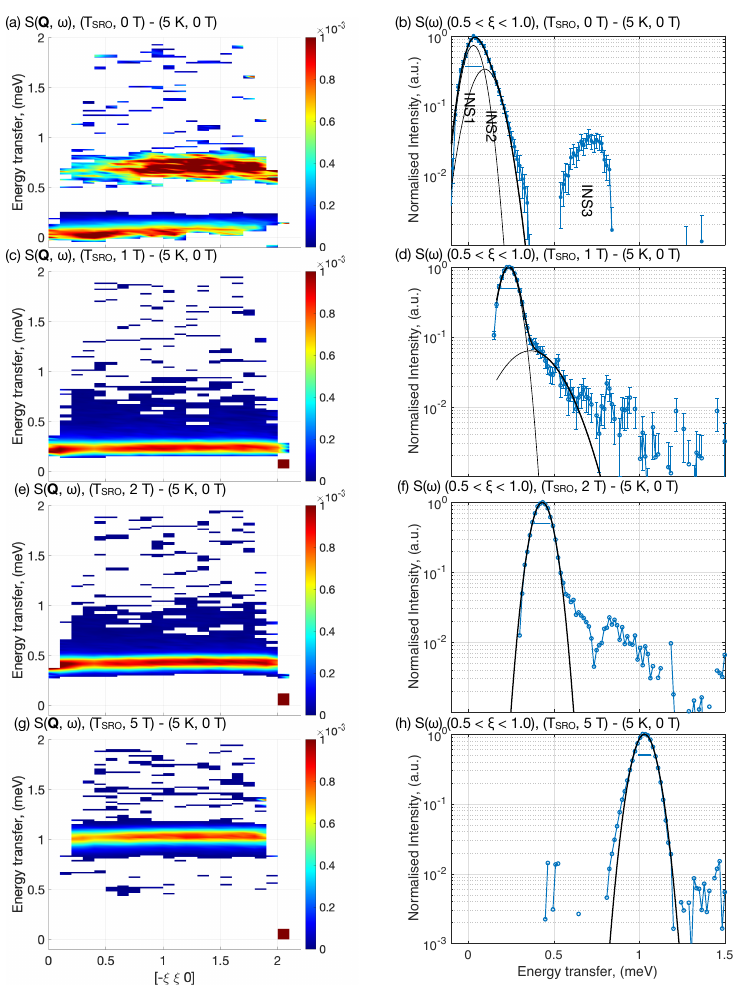}
		\caption{ (a,c,e,g) Magnetic field dependence (logarithmic colorscale) of S([-$\xi$ $\xi$ 0], $\omega$) ($E_{i}$ = 3.27 meV, data integrated across  -0.2 $\leq$ $\zeta$ $\leq$ 0.2 in [$\zeta$, $\zeta$, -2$\zeta$] and background subtracted) at $T_{SRO}$ for (a) $\mu_{0} H$ = 0 T. Any intensity above $\Delta$E = 0.345 meV increased by a factor of 20. (c) $\mu_{0} H$ = 1 T,  (e) $\mu_{0} H$ = 2 T and (g) $\mu_{0} H$ = 5 T. (b,d,f,h) Magnetic field dependence (logarithmic colorscale) of $S(\textbf{Q},\omega)$, Q integrated across 0.5 $<$ $\xi$ $<$ 1.0, for $\mu_{0} H$ = 0, 1, 2 and 5 T. Continuous lines are Gaussian fits. The horizontal line represents the instrumental energy resolution, $\delta E$, at the energy transfer shown.}
        \label{fig:5A_FieldDep}
\end{figure}

An overview of the field dependence of $S(\textbf{Q}, \omega)$ at $T_{SRO}$ is presented in Fig. \ref{fig:5A_FieldDep} (a,c,e,g) for $E_{i}$ = 3.27 meV with $\mu_{0} H$ = 0, 1, 2, and 5 T. Data shown in Figs. \ref{fig:5A_FieldDep} are extracted from the 3D data sets integrated across  -0.2 $\leq$ $\zeta$ $\leq$ 0.2 in [$\zeta$, $\zeta$, -2$\zeta$]. 





The magnetic contribution was isolated by subtracting the instrumental background and nuclear contributions of $S(\textbf{Q}, \omega)$ measured at $\mu_{0} H$ = 0 T and 5 K. A cut through the magnetic excitations, $S(\omega)$ integrated across 0.5 $<$ $\xi$ $<$ 1.0, is shown in Figs. \ref{fig:5A_FieldDep} (b,d,f,h) to present clearly the field dependence of the peak positions, linewidths and any further broad scattering features.  

Fig. \ref{fig:5A_FieldDep}(a,b), $S(\textbf{Q}, \omega)$ and $S(\omega)$ for $\mu_{0}H$ = 0 T,  shows an inelastic nearly dispersionless contribution at $\Delta E$ = 0.02(1) and 0.096(2) meV (INS1,INS2) and a further nearly dispersionless excitation at 0.7 meV (INS3), consistent with previous results \cite{YbGG_Lise_2021_PhysRevB.104.064425,lHotel2021PhysRevB.104.024427}. In Fig. \ref{fig:5A_FieldDep}(a) any intensity above $\Delta E$ = 0.345 meV is increased by a factor of 20 to provide a clear overview of the low and higher energy excitations simultaneously. Upon applying $\mu_{0} H$ = 1 T, Fig.\ref{fig:5A_FieldDep}(c,d), we observe a single broad dispersionless excitation centered on $\Delta E$ = 0.237(2) meV and a broad feature extending to approximately $\Delta E$ = 1 meV, see Fig. \ref{fig:5A_FieldDep}(d), a portion of which we can model with a broader Gaussian lineshape. It is not possible to clearly separate INS1, 2 and 3.  In the elastic line there is only a strong Bragg peak at [-2 2 0] consistent with a ferromagnetic alignment of spins. The magnetic field dependence of this Bragg peak is shown in the supplementary materials \ref{sec:Supp}, Fig. \ref{fig:II_220}. 
Upon further increasing the applied magnetic field, $\mu_{0} H$ = 2 T, we observe a single main excitation that is well represented with a Gaussian lineshape broader than the instrumental energy resolution Fig. \ref{fig:5A_FieldDep}(e,f). Similar to the scattering for $\mu_{0} H$ = 1 T, broad scattering, up to 1.2 meV, extends beyond the main excitation. The main excitation continues to increase in energy transfer for $\mu_{0} H$ = 5 T with a further broadening of the linewidth. The broad scattering at higher energy transfer, a feature for both  $\mu_{0} H$ = 1 and 2 T, is no longer observed for $\mu_{0} H$ = 5 T, see Fig. \ref{fig:5A_FieldDep}(g,h). 

\begin{figure}[!htp]
    	\centering
    	\includegraphics[width =0.5\linewidth]{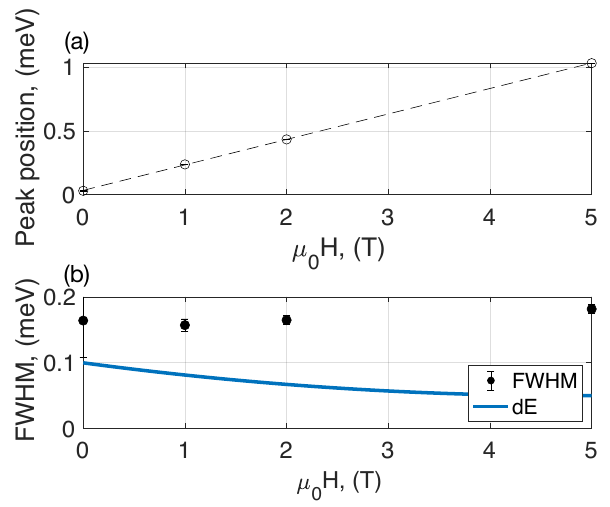}
		\caption{(a) Magnetic field dependence of the peak position following a linear dependence  (dashed line)(b) magnetic field dependence of the peak width (FWHM). Continuous line shows corresponding instrumental energy resolution $\delta$E.}
        \label{fig:HDep_Pos_FWHM}
\end{figure}
Fig. \ref{fig:HDep_Pos_FWHM} shows the magnetic field dependence of (a) the peak position and (b) the Full Width at Half Maximum (FWHM) of the main magnetic excitation, observed in Figs. \ref{fig:5A_FieldDep}. We demonstrate the Zeeman effect previously reported by Lhotel {\it et} al. \cite{lHotel2021PhysRevB.104.024427}. The slope of the magnetic field dependence is 0.1953(3) meV/T and the 0 T with an intercept of 0.029(1) meV, see Fig. \ref{fig:HDep_Pos_FWHM}(a), again consistent with the values previously obtained.

 The FWHM of the excitation is broader than the instrumental resolutions, continuous line, Fig. \ref{fig:HDep_Pos_FWHM}(b) for all applied magnetic fields and, furthermore, remains broader than that expected from linear spin wave theory (LSWT) in the ordered regime for $\mu_{0} H$ = 5 T , see section \ref{LSWT}. 

\begin{figure}[!htp]
    	\centering
    \includegraphics[width =\linewidth]{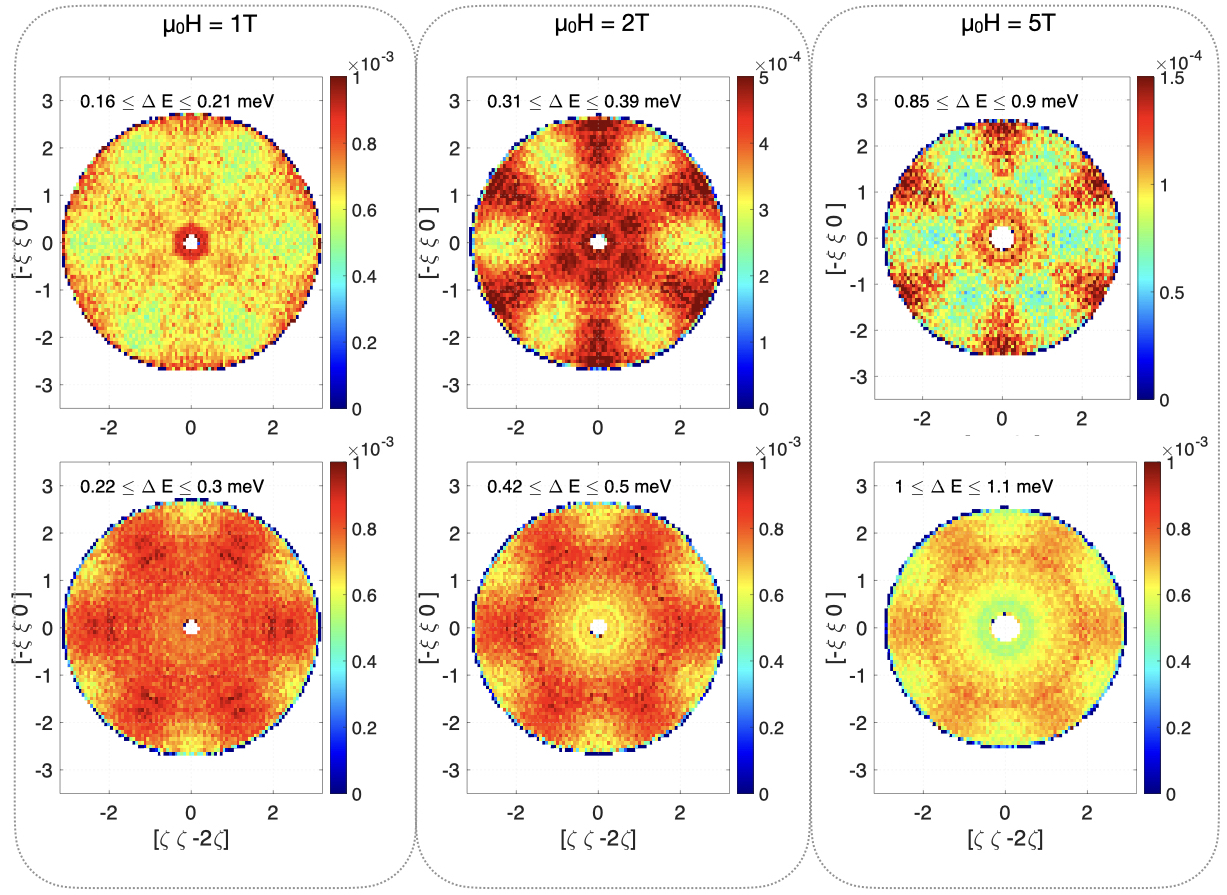}
        \caption{ $S({\textbf Q})$ dependence of flat band mode at $T_{SRO}$,  integrated in energy across lower and higher regions, for (a,b) $\mu_{0}H$ = 1 T (a) 0.13 $\leq$ $\Delta E$ $\leq$ 0.21 meV,  (b) 0.22 $\leq$ $\Delta E$ $\leq$ 0.30 meV, (c,d) for $\mu_{0} H$ = 2 T (c) 0.31 $\leq$ $\Delta E$ $\leq$ 0.39 meV, (d) 0.42 $\leq$ $\Delta E$ $\leq$ 0.5 meV  and (e,f) $\mu_{0}H$ = 5 T (e) 0.85 $\leq$ $\Delta E$ $\leq$ 0.9 meV and (f) 1 $\leq$ $\Delta E$ $\leq$ 1.1 meV.}
        \label{fig:SQQ_1_2_5T}
\end{figure}

 The \textbf{Q} dependence of the main magnetic mode is presented in Figs. \ref{fig:SQQ_1_2_5T} for $\mu_{0} H$ = 1, 2 T and 5 T, (a,b), (c,d), (e,f) respectively. The dispersive nature of the mode is observed by extracting a low and high energy region, for example for $\mu_{0} H$ = 1 T,  integrating across 0.13 $<$ $\Delta E$ $<$ 0.21 meV, Fig. \ref{fig:SQQ_1_2_5T}(a), for the bottom of the dispersion and across 0.22 $<$ $\Delta E$ $<$ 0.3 meV, Fig. \ref{fig:SQQ_1_2_5T}(b), for the top of the dispersion. For all applied magnetic fields there are broad but very distinct features that follow the crystalline 6-fold symmetry in both the lower and higher energy regions. The low energy  features,  (a, c, e), are mostly inverted with respect to the high energy features, (b, d, f) for $\mu_{0} H$ = 1, 2 and 5 T. First, for $\mu_{0} H$ = 1 T, we number the main features for the low energy regions, in Fig. \ref{fig:SQQ_1_2_5T}(a), (1-4) with (1) a low $\textbf{Q}$ high intensity hexagonal scattering feature, up to $\xi$ $\sim$ 0.25 along [-$\xi$ $\xi$ 0], that is separated by (2) a weaker scattering area from (3) a broader band of scattering, starting at $\xi$ $>$ 0.5 for [-$\xi$ $\xi$ 0], before extending in a triangular feature (3) juxtaposed to a weaker scattering region (4). Upon increasing the magnetic field to $\mu_{0} H$ = 2 T, region (2) is more structured with a distinct intense region for 0.5 $<$ $\xi$ $<$ 1 along [-$\xi$ $\xi$ 0], then a reduced intensity region 1.0 $<$ $\xi$ $<$ 1.5 along [-$\xi$ $\xi$ 0] resulting in an intense triangular region starting at 1.5 along [-$\xi$ $\xi$ 0]. Increasing the magnetic field to $\mu_{0} H$ = 5 T extends the low $\textbf{Q}$ region further and reveals a clear separation in intensity between region (1) and region (3).

\newpage
\subsection{High resolution measurement of low-lying excitations for $\mu_{0}H$  = 0 and 0.5 T.}

\begin{figure}[!htp]
        \centering
    \includegraphics[width =\linewidth]{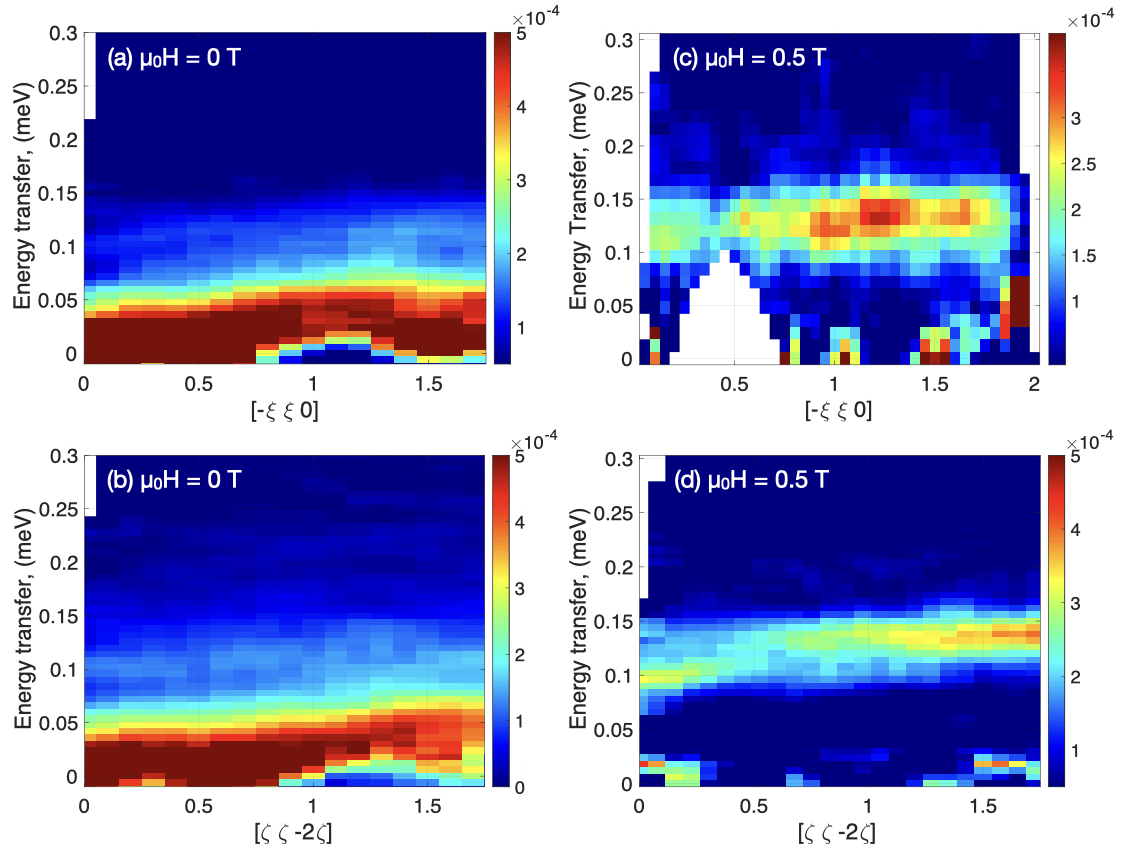}
	\caption{Magnetic field dependence, at T = $T_{SRO}$, of S( [-$\xi$ $\xi$ 0], $\omega$)) and S( [$\zeta$ $\zeta$ -2$\zeta$], $\omega$)) ($E_{i}$ = 1.67 meV, background subtracted) at $T_{SRO}$ for (a,b) $\mu_{0}H$ = 0 T,(c,d) $\mu_{0}H$ = 0.5 T. Data are extracted from the 3D data sets integrated across  -0.2 $\leq$ $\zeta$ $\leq$ 0.2 in [$\zeta$, $\zeta$, -2$\zeta$] for (a) and (c) and -0.1 $\leq$ $\zeta$ $\leq$ 0.1 in [$-\xi$, $\xi$, 0] for (b) and (d).}
\label{SQW_7AH}
\end{figure} 


The lowest lying magnetic excitations, for $T_{SRO}$, in both zero magnetic field and under $\mu_{0}H$ = 0.5 T, Figs. \ref{SQW_7AH}, are further examined with higher energy resolution using $E_{i}$ = 1.67 meV with $\Delta E$ = 0.03 meV. Data shown in Figs. \ref{SQW_7AH} are extracted from the 3D data sets integrated across  -0.2 $\leq$ $\zeta$ $\leq$ 0.2 in [$\zeta$, $\zeta$, -2$\zeta$] for (a) and (c) and -0.1 $\leq$ $\zeta$ $\leq$ 0.1 in [$-\xi$, $\xi$, 0] for (b) and (d). 
The low lying magnetic excitations were isolated by removing a background contribution measured with $\mu_{0}H$ = 5 T and at $T$ = $T_{SRO}$ which shows no low lying contributions, see Fig. \ref{fig:5A_FieldDep}(d). 

Under $\mu_{0}H$ = 0.0 T a weak dispersion is observed for both crystalline planes with zone boundaries at [-0.5 0.5 0] and [-1.5 1.5 0], Fig. \ref{SQW_7AH}(a), and  [0.5 0.5 -1.0] and [1.5 1.5 -3], Fig. \ref{SQW_7AH}(b). The zone boundaries of the dispersions along [-$\xi$ $\xi$ 0] are consistent with the recent determination of the magnetic propagation vector found for YbGG \cite{Raymond2024}, $\textbf{k}$ = ($\frac{1}{2}$ $\frac{1}{2}$ 0), which would indicate either that we are in the ordered state below T$_{N}$, an unlikely scenario, or that there are some low lying excitations reminiscent of the ordered phase coexisting with a disordered phase, a possible spin slush state. A spin slush state exhibits extremely slow relaxation for some spins while others fluctuate quickly down to zero temperature and has been suggested for GGG \cite{Gingras2016}. Above the correlated magnetic dispersive mode there is significant extra scattering extending up to 0.2 meV for both crystalline directions substantiating the possibility of a spin slush phase.  
Upon the application of $\mu_{0}H$ = 0.5 T gapped flat magnetic excitations are lifted in energy, see Fig. \ref{SQW_7AH}(c,d), with a slight softening of the mode for low $\textbf{Q}$ regions. The continuum scattering observed above the dispersive mode for $\mu_{0}H$ = 0.0 T is not visible for  $\mu_{0}H$ = 0.5 T.

 \begin{figure}[!htp]
    	\centering
    	\includegraphics[width =0.5\linewidth]{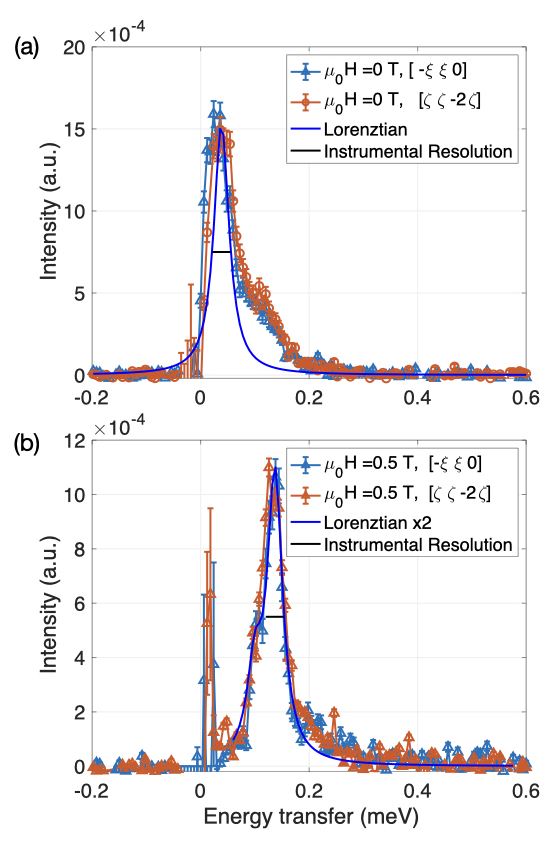}
		\caption{Line cut, derived from Fig. \ref{SQW_7AH} from $\Delta$E =0.0 extending up to $\Delta$E =0.4 meV along the two principal crystalline directions [-$\xi$ $\xi$ 0] and [$\zeta$ $\zeta$ -$2\zeta$ ], integrated at the peak of the dispersion curves,  1 $\leq$ $\xi$ $\leq$ 1.2 and  1 $\leq$ $\zeta$ $\leq$ 1.2, under applied magnetic fields $\mu_{0}H$ = 0 and 0.5 T, at $T$ = $T_{SRO}$.}
        \label{7A_LinePlotFieldDep}
\end{figure}

Figs. \ref{7A_LinePlotFieldDep} shows \textbf{Q} integrated linecuts from Figs. \ref{SQW_7AH}, at the top of the dispersion curve along the two principal crystalline directions probed, [-$\xi$ $\xi$ 0] and [$\zeta$ $\zeta$ -$2\zeta$ ], integrated across 1.0 $\leq$ $\xi$ $\leq$ 1.2  and 1.0 $\leq$ $\zeta$ $\leq$ 1.2, under applied magnetic fields (a) $\mu_{0}H$ = 0 T and (b) 0.5 T. For $\mu_{0}H$ = 0 T two low-lying non energy resolved excitations are  observed at $\Delta E$ = 0.037(3) meV, a principal mode, and at $\Delta E$ = 0.113(2) meV, a broader region, for both direction [-$\xi$ $\xi$ 0] and [$\zeta$ $\zeta$ -$2\zeta$ ], Fig. \ref{7A_LinePlotFieldDep}(a). A Lorentzian lineshape, consistent with the instrumental resolution, is provided as a guide to the eye for the principal excitation. The slight broadening of this principal mode, relative to the instrumental resolution, is due to the dispersion. The higher energy mode, 0.1 $\leq$ $\Delta E$ $\leq$ 0.2 meV, is significantly broader than the instrumental resolution.  Upon the application of  $\mu_{0}H$ = 0.5 T the principal mode, also equivalent for both directions, increases in energy transfer, Fig. \ref{7A_LinePlotFieldDep}(b). The continuous line in Fig. \ref{7A_LinePlotFieldDep}(b) for $\mu_{0}H$ = 0.5 T is a two Lorentzian fit to the gapped excitation centered at $\Delta E$ = 0.090(5) and $\Delta E$ = 0.137(5) meV with a linewidth equal to the instrumental resolution. The double mode can again be assigned to the dispersive nature of this excitation. The broad continuum, however, is not observed, although there remains a weak contribution above the instrumental resolution. 

\begin{figure}[!htp]
    	\centering
   \includegraphics[width =0.8\linewidth]{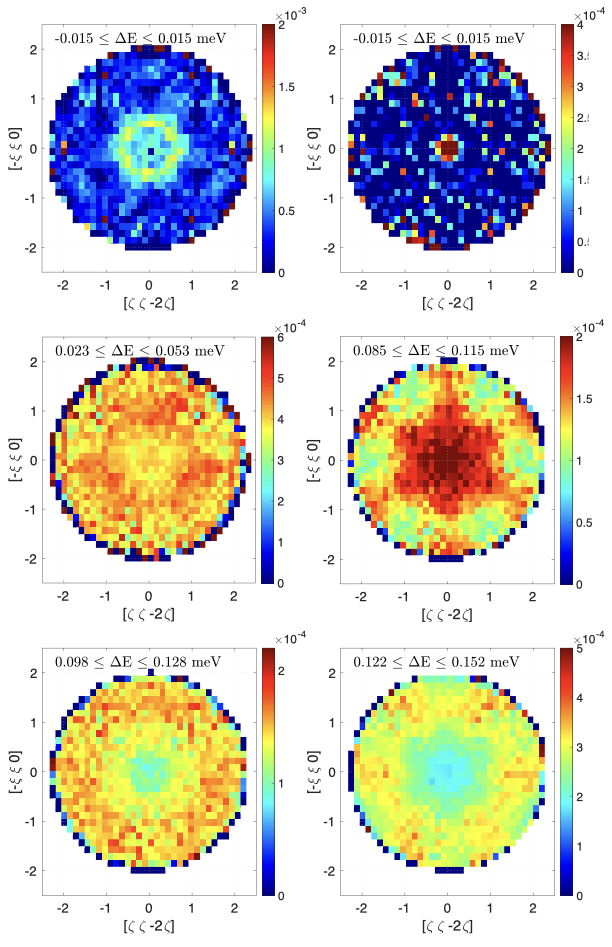}
		\caption{S(\textbf{Q}) at $T$ = $T_{SRO}$, $E_{i}$ = 1.67 meV, (a-c) $\mu_{0}H$ = 0.0 T and (d-f) $\mu_{0}H$ = 0.5 T. Energy transfer centered on (a, d) $\Delta E$=0.0 meV,(b, e) $\Delta E$= low energy modes,(c, f) $\Delta E$= high energy modes, as outlined in text.}
        \label{7A_SQQFieldDep_0T}
\end{figure}

The {\textbf Q} dependence can again be accessed via a cut through the energy landscape of Figs. \ref{SQW_7AH} with an energy width equal to the instrumental resolution.  Figs. \ref{7A_SQQFieldDep_0T} therefore provide the elastic S(\textbf{Q}) for $\mu_{0}H$ = 0 (a-c) and for $\mu_{0}H$ = 0.5 T (d-f). 
The \textbf{Q} dependence of the elastic scattering for $\mu_{0}H$ = 0 T , see Fig. \ref{7A_SQQFieldDep_0T}(a), reveals magnetic scattering that is elliptical in \textbf{Q} space. 
This scattering is derived from magnetic structures with spatial scales ranging between 14(2) - 17(1) \AA{}. Upon applying $\mu_{0}H$ = 0.5 T, Fig. \ref{7A_SQQFieldDep_0T}(d), the elliptical magnetic scattering is reduced in reciprocal space with only very low $\textbf{Q}$ scattering remaining that can be assigned to magnetic structures with spatial scales ranging between 61(2) and 86(3) \AA{}. In the inelastic channels, for $\mu_{0}H$ = 0.0 T, Fig. \ref{7A_SQQFieldDep_0T}(b-c), both the low and high energy modes, 0.023 $<$ $\Delta E$ $<$ 0.053 meV and 0.098 $<$ $\Delta E$ $<$ 0.128 meV, respectively, show a broad band of scattering away from $\textbf{Q}$ = 0 with some structure, albeit difficult to distinguish. Under $\mu_{0}H$ = 0.5 T the magnetic scattering shows clear structure for the low lying energy mode, 0.085 $<$ $\Delta E$ $<$ 0.115 meV , Fig. \ref{7A_SQQFieldDep_0T}(e), with significant scattering at low $\textbf{Q}$ which then extends into arms of scattering with 6 fold symmetry, very similar to the dispersive scattering observed at $\mu_{0}H$ = 1-5 T, see Figs. \ref{fig:SQQ_1_2_5T}, that is similarly dispersive with a  $\textbf{Q}$ dependence of the higher energy mode, 0.122 $<$ $\Delta E$ $<$ 0.158 meV, that opposes those of the low energy mode. The similarities of the $S({\textbf Q})$ profiles of the low and high energy modes as the applied magnetic field is increased to $\mu_{0}H$ = 5 T, see Figs. \ref{fig:SQQ_1_2_5T}, indicates the robust nature of these excitations.

\newpage
\subsection{Linear Spin Wave Theory}
\label{LSWT}

\begin{figure}[!htp]
    	\centering
    	\includegraphics[width = \linewidth]{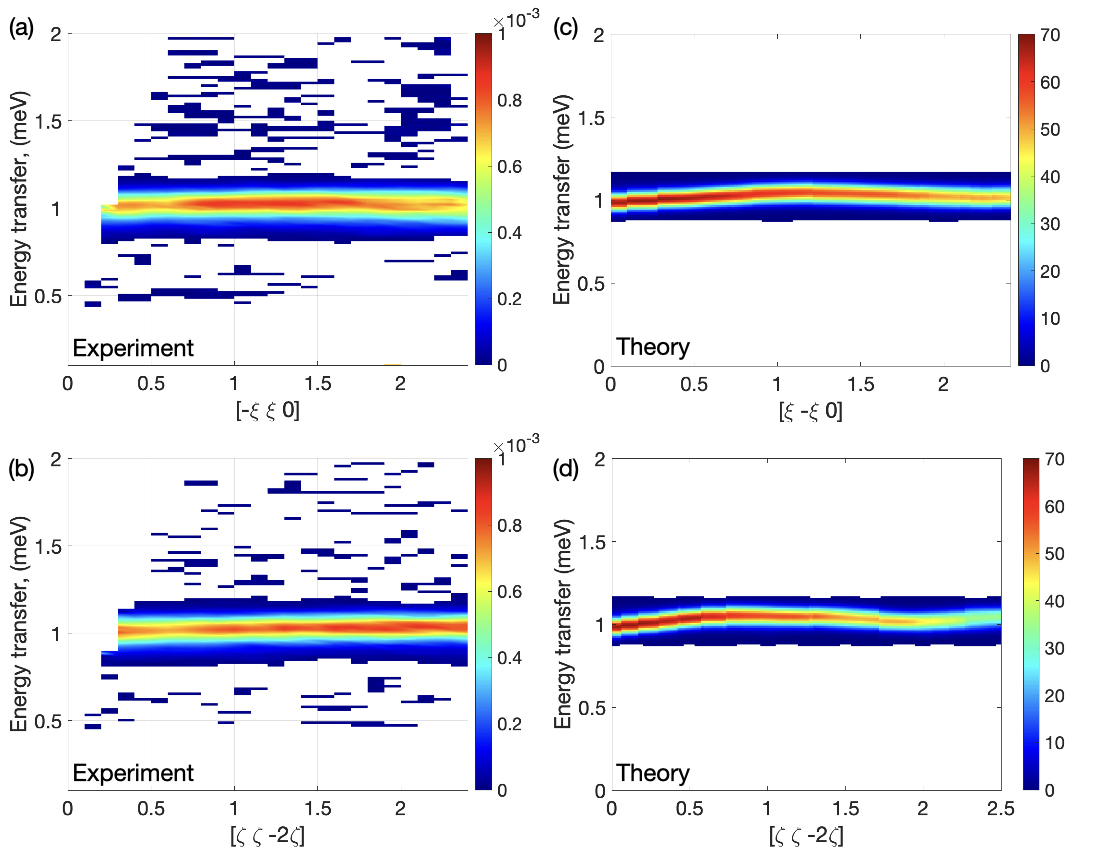}
		\caption{(a,b) Experimental $S(\textbf{Q}, \omega)$ under applied field $\mu_{0}H$ = 5 T at $T_{SRO}$ for $\textbf{Q}$ =  [-$\xi$ $\xi$ 0] and  [$\zeta$ $\zeta$ -2$\zeta$], respectively.(c), (d) corresponding calculated $S(\textbf{Q}, \omega)$ as described in the text.}
        \label{fig:YbGG_SQW_5T_TheoryExp}
\end{figure}

\begin{figure}[!htp]
    	\centering
        \includegraphics[width = \linewidth]{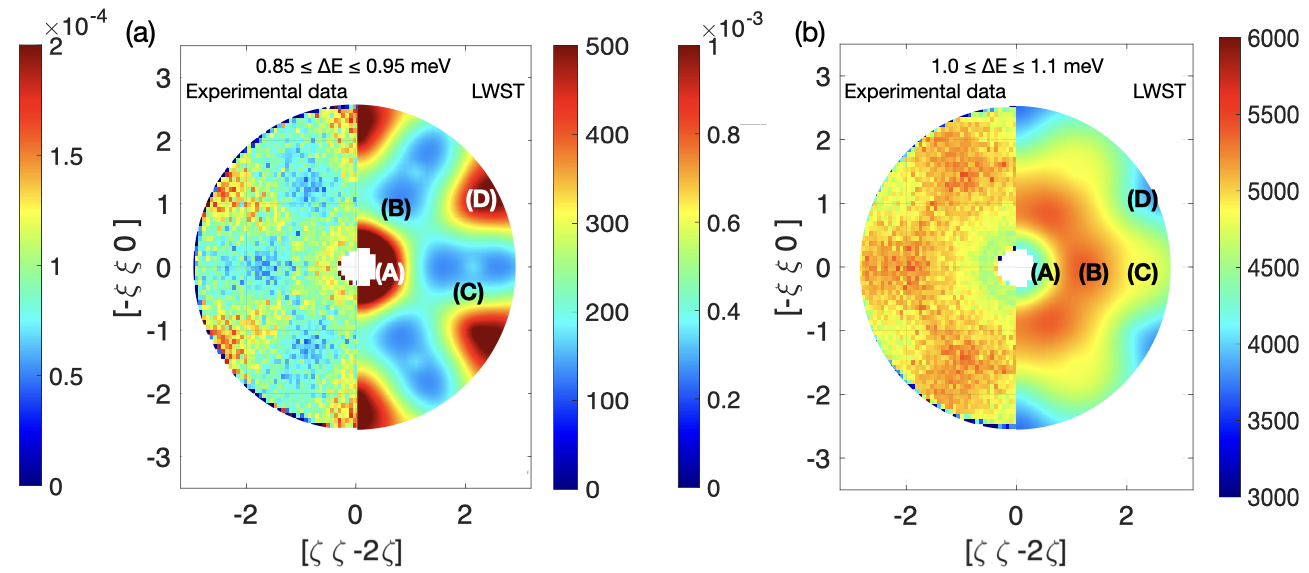}
                \caption{(a) Experimental S(\textbf{Q}) under applied field $\mu_{0}H$ = 5 T at $T_{SRO}$ for the (a) low energy mode, 0.85 $\leq$ $\Delta$E $\leq$ 0.95 meV and (b) high energy mode 1.0 $\leq$ $\Delta$E $\leq$ 1.1 meV. Figs. Right hand side of each figure is the representative LSWT as described in text.}
        \label{fig:YbGG_SQQ_5T_TheoryExp}
\end{figure}

Linear spin wave theory (LSWT) provides the principal term of $S(\textbf{Q}, \omega)$ for long-range ordered magnetic systems and allows us to extract the classical features of this data. We employ the linear spin wave package, SpinW, to calculate  $S(\textbf{Q}, \omega)$ for the ferromagnetically ordered state with $\mu_{0}H$ = 5 T along the [1 1 1] crystallographic direction \cite{SpinW}. The calculation, inclusive of magnetic form factor, $g$-factor and instrumental resolution, that provides the most consistency between experiment and theoretical $S(\textbf{Q}, \omega)$ and S(\textbf{Q}) at various energy transfers is found with a dipole exchange interaction, $D$ = 101 mK, in accordance with Lhotel {\it et} al. extended out to 30 \AA{}, and $J_{1}$ = $J_{3}$ = -15 mK. Both the anisotropic value $g$ = [2.84, 3.59, -3.72], determined by Carson and Sandberg \cite{Carson1960,YbGG_Lise_2021_PhysRevB.104.064425}, and the isotropic value $g$ = 3.42, considered by Lhotel \cite{lHotel2021PhysRevB.104.024427}, were considered but do not provide any variation in $S(\textbf{Q}, \omega)$ within the instrumental energy and \textbf{Q} resolution. The Zeeman energy provides the energy scale of the excitation, see Fig. \ref{fig:YbGG_SQW_5T_TheoryExp},  with $D$, $J_{1}$ and $J_{3}$ providing the structure observed in $S(\textbf{Q})$, see Fig. \ref{fig:YbGG_SQQ_5T_TheoryExp}. An overview of $S(\textbf{Q}, \omega)$ and $S(\textbf{Q})$ for a range of other exchange interactions is given in the supplementary materials section, section \ref{sec:Supp}, to justify the exchange interactions we estimate. 

First, we compare the measured $S(\textbf{Q}, \omega)$ along the two crystalline directions Fig. \ref{fig:YbGG_SQW_5T_TheoryExp}(a,b) to the model, Fig. \ref{fig:YbGG_SQW_5T_TheoryExp}(c,d) . LSWT is able to reproduce a flat band at the correct energy transfer since this is principally derived from the Zeeman energy. There are slight differences that include a very weak high $\textbf{Q}$ dispersion in the calculated $S(\textbf{Q}, \omega)$ and a linewidth that is narrower than the experimental result. The experimentally deduced FWHM of the excitation is 0.192(2) meV for $S(\textbf{Q})$ integrated across 0.5 $<$ $\xi$ $<$ 1.0, from Fig. \ref{fig:5A_FieldDep}(h), while LSWT simulates a FWHM of 0.153 meV for an equivalent integration of reciprocal space. 

Second, we consider the {\bf Q} dependence of the excitation with low and high energy modes. Fig. \ref{fig:YbGG_SQQ_5T_TheoryExp} compares the experimental (a,b) and calculated (c,d) $S(\textbf{Q}, \omega)$ for $\mu_{0}H$ = 5 T and $T_{SRO}$ along the two principal crystalline axis [-$\xi$ $\xi$ 0] and  [$\zeta$ $\zeta$ -2$\zeta$] for (a,c) the low energy region 0.85 $<$ $\Delta E$ $<$ 0.95 meV and (b,d) the high energy region 0.1 $<$ $\Delta E$ $<$ 1.1 meV. 

Again, the salient features are comparable with a strong six-fold symmetry for both energy modes. First, for the low energy mode, see Figs. \ref{fig:YbGG_SQQ_5T_TheoryExp}(a,c), we are able to theoretically reproduce the low $\textbf{Q}$ region (A). As we move towards higher $\textbf{Q}$, there is experimentally a weaker scattering region (B) extending into the triangular six fold arms (D) features that are well reproduced by LSWT. A weaker region in between the two triangular (D) shapes shows, in the calculation, a linear dissecting feature, (C), also observed in the experimental signature, however it is rather weak. 


In the high energy region, Figs.\ref{fig:YbGG_SQQ_5T_TheoryExp}(b,d), the features observed in the lower energy mode are inverted. A low intensity scattering region is observed experimentally for $\textbf{Q}$ $\leq$ 0.5 \AA$^{-1}$ and reproduced theoretically (A). Furthermore, the broad band of scattering (B) extends into arms of scattering with six fold symmetry (C) separated by weak scattering regions (D), observed in both the experimental and theoretical signatures. 



\section{Discussion and conclusion}

We have conducted a detailed inelastic neutron scattering investigation of the magnetic field dependence of the magnetic excitations in single crystal YbGG in the short ranged order temperature regime $T_{SRO}$ $\sim$ 200 mK with $\mu_{0}H$ aligned with the [1 1 1] crystallographic axis. The magnetic excitations have been accessed across a broad region of reciprocal space providing a detailed overview of the energy and momentum landscape. 

In zero applied field, the diffraction profile is consistent with scattering from nano-sized magnetic structures, with a spatial scale of a 10 spin loop while very low energy correlated magnetic spin waves are observed 
that show some consistency with the recently determined long range order in YbGG \cite{Raymond2024}. However, continuum scattering exists above the correlated features for $\mu_{0}H$ = 0 T suggestive of a spin slush phase previously theoretically proposed for GGG \cite{Rau_2016} or the signatures of quantum fluctuations \cite{QuantumFluctuationsJinLong_2024}.

Upon applying a weak magnetic field,  $\mu_{0}H$ = 0.5 T,  the magnetic structure, observed in the diffraction profile, extends in size, while the inelastic scattering reveals mostly flat, soft modes, with a {\textbf Q} dependence that shows both short and longer ranged dispersive correlations. These modes reveal a \textbf{Q} dependence that is reminiscent of the excitations in Nd$_{2}$Zr$_{2}$O$_{7}$ in which the flat mode is a feature of the Coulomb spin liquid phase, suppressed by the onset of antiferromagnetic ordering \cite{Petit_2016_Spin_Ice}. 
In the case of YbGG, for $\mu_{0}H$ = 0.5 T, large magnetic structures coexist with a fluctuating state. The continuum scattering is not visible for $\mu_{0}H$ = 0.5 T. 
As we increase applying the magnetic field to $\mu_{0}H$ = 1 T we only observe ferromagnetic scattering as the ordered component, a soft mode increases in energy consistent with a Zeeman dependence and reveals a similar dispersive \textbf{Q} dependence, as first observed for $\mu_{0}H$ = 0.5 T. It is of interest to note that under $\mu_{0}H$ = 1 T significant continuum scattering above the soft modes is again observed. 

The $\textbf{Q}$ dependence of these energy modes show only a subtle change upon further application of magnetic field yet with the position of these excitations, in energy transfer, continuing to follow a linear Zeeman dependence. Continuum scattering, above the soft modes in energy, is observed for 1 $\leq$ $\mu_{0}H$ $\leq$ 2 T, but disappears in the magnetically saturated regime, $\mu_{0}H$ = 5 T.  

LSWT is employed to determine the exchange interactions in the magnetically saturated region for $\mu_{0}H$ = 5 T. We subdivide the magnetic excitations into two distinct energy regimes with a low and high energy region that reveals clear modulated, in 
${\textbf Q}$, structures. The $\textbf{Q}$ dependence of these distinct energy regions is well described by employing a dominant dipolar exchange interaction, $D$ = 101 mK, and much weaker near neighbour and inter-hyperkagome exchange antiferromagnetic interactions, $J_{1}$ = $J_{3}$ = -15 mK. YbGG can therefore be described as a quantum dipolar magnet with highly robust soft modes. Further studies will be needed to understand the details of the ground state of YbGG, provide a theoretical understanding of the continuum scattering observed, in addition to lifetime broadening of the excitations beyond the dispersion widths expected from classical calculations. We expect that these results will stimulate theoretical considerations and in particular lead to the development of models to describe these complex magnetic states. 

\section{Acknowledgments}
The authors would like to acknowledge the excellent sample environment and data analysis suppport of the Institut Laue Langevin. The neutron scattering data collected at the ILL for the present work are available at ref. \cite{IN5Data_2021}. PD and KL would like to acknowledge the support provided by the Danish Committee for Research infrastructure (NUFI) through DanScatt and the Danish Lighthouse QMat. ER and EL acknowledge financial support from Agence Nationale de la Recherche, France, Grant No. ANR-18-CE05-0023. The work at the University of Warwick was supported by EPSRC, UK through Grant EP/T005963/1.

\section{Supplementary Materials}
\label{sec:Supp}
\section{Magnetic field dependence of [-2 2 0]}
\begin{figure}[!htp]
    	\centering
   \includegraphics[width =0.5\linewidth]{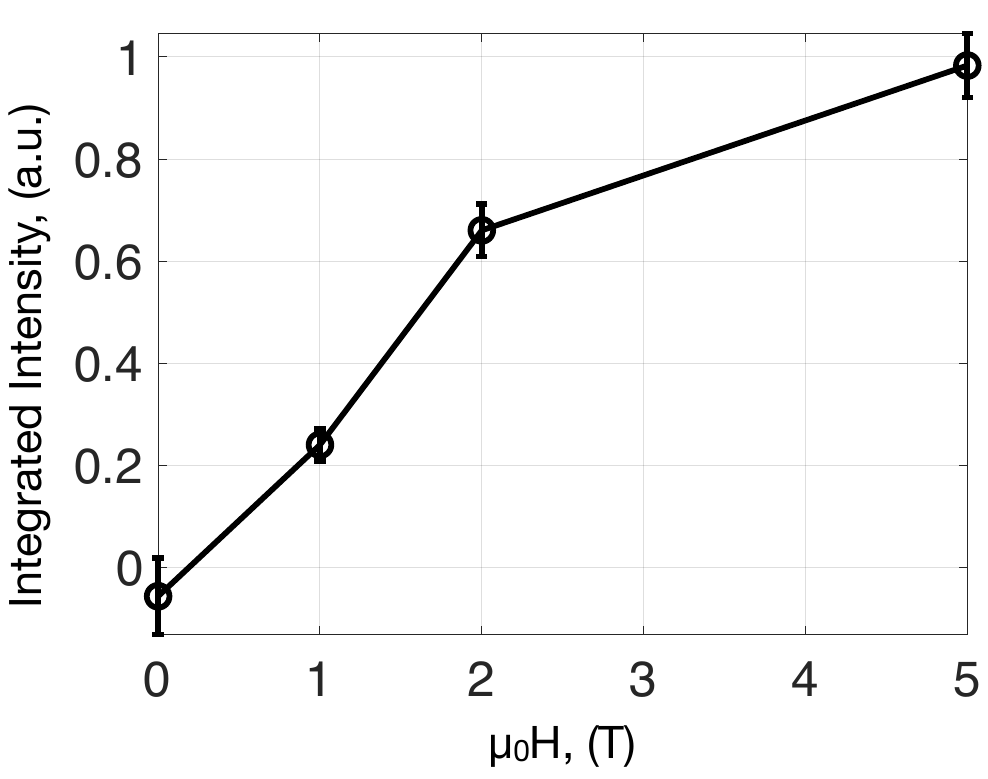}
		\caption{Magnetic field dependence of [-2 2 0] Bragg peak position.}
        \label{fig:II_220}
\end{figure}

The magnetic field dependence of the magnetic Bragg peak upon the application of a magnetic field is shown in Fig. \ref{fig:II_220}, integrated across the elastic line from Fig \ref{fig:5A_FieldDep} and for 1.8 $\leq$ $\xi$ $\leq$ 2.1 along [$-\xi$ $\xi$ 0].  

\section{Linear spin wave theory for a range of exchange interactions}
An overview of some $S(\textbf{Q}, \omega)$ and $S(\textbf{Q}$, $\textbf{Q})$ profiles for various exchange interactions, calculated as described in the main text but with the exchange interactions indicated in the label.  

\begin{figure}[!htp]
    	\centering
   \includegraphics[width =\linewidth]{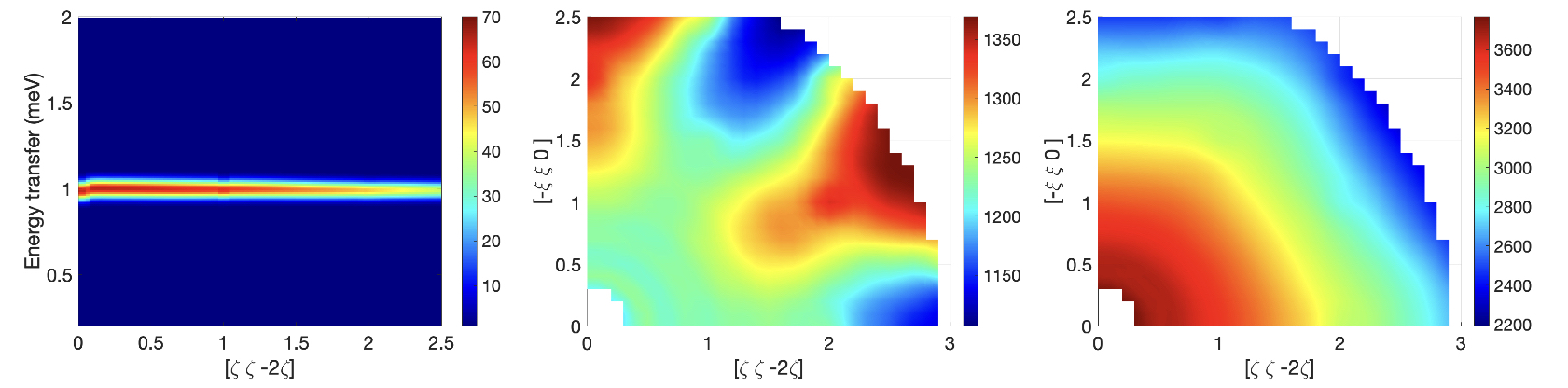}
		\caption{LSWT S(\textbf{Q}, $\omega$) and S(\textbf{Q}) profiles with $D$ = 110 mK only (middle) low energy mode, 0.85 $\leq$ $\Delta$E $\leq$ 0.95 meV and (right) high energy mode 1.0 $\leq$ $\Delta$E $\leq$ 1.1 meV.}
        \label{fig:SQW_SQQ_Donly}
\end{figure} 

\begin{figure}[!htp]
    	\centering
   \includegraphics[width =\linewidth]{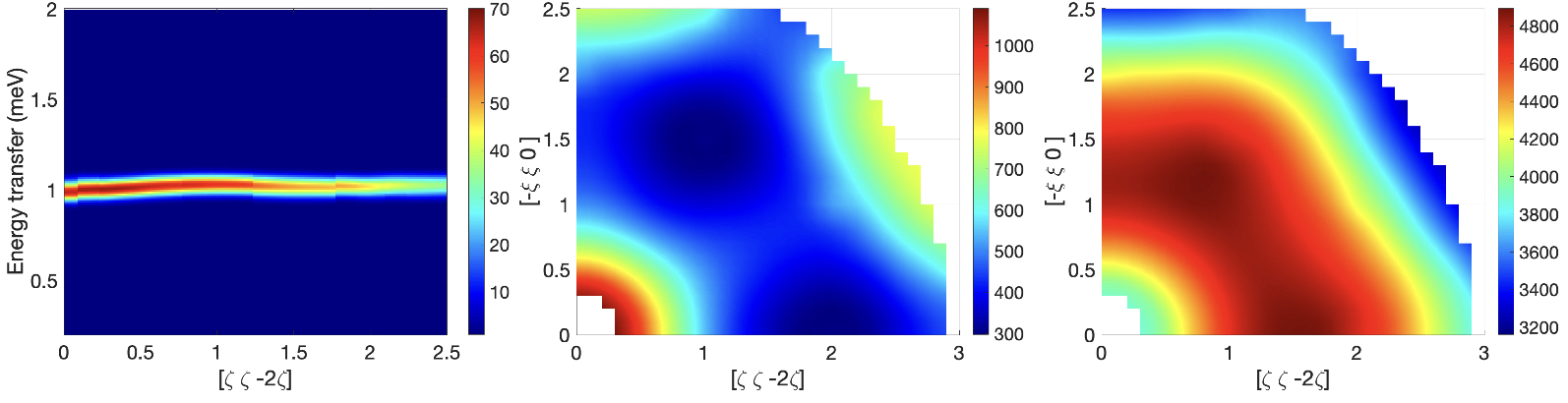}
		\caption{LSWT S(\textbf{Q}, $\omega$) and S(\textbf{Q}) profiles with $D$ = 110 mK and $J_{1}$ = -15 mK, (middle) low energy mode, 0.85 $\leq$ $\Delta$E $\leq$ 0.95 meV and (right) high energy mode 1.0 $\leq$ $\Delta$E $\leq$ 1.1 meV.  }
        \label{fig:SQW_SQQ_DJ1only}
\end{figure} 

\begin{figure}[!htp]
    	\centering
   \includegraphics[width =\linewidth]{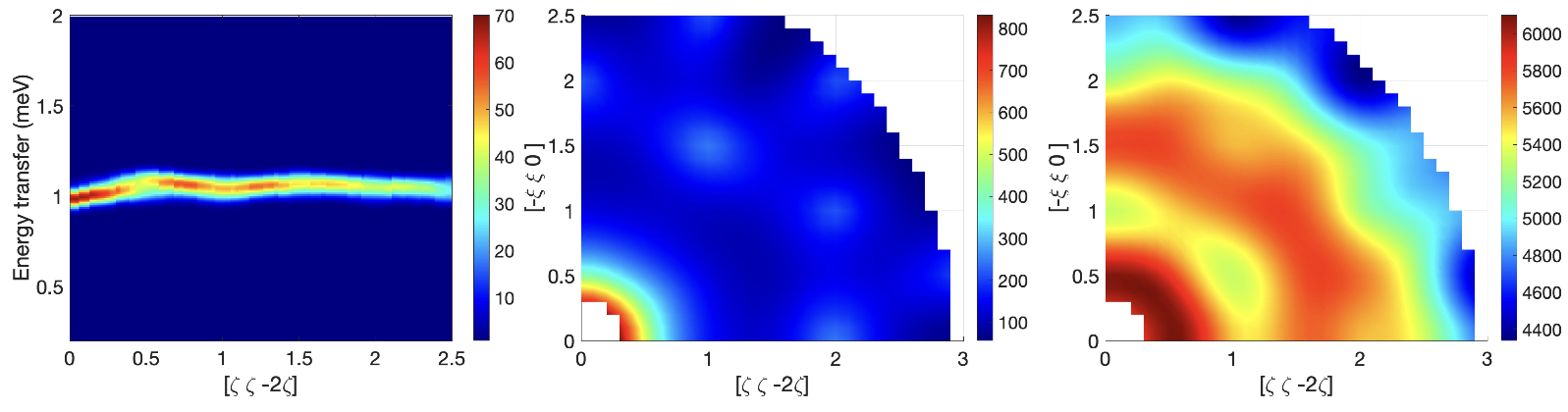}
		\caption{LSWT S(\textbf{Q}, $\omega$) and S(\textbf{Q}) profiles with $D$ = 110 mK and $J_{2}$ = -15 mK, (middle) low energy mode, 0.85 $\leq$ $\Delta$E $\leq$ 0.95 meV and (right) high energy mode 1.0 $\leq$ $\Delta$E $\leq$ 1.1 meV.}
        \label{fig:SQW_SQQ_DJ2only}
\end{figure} 

\begin{figure}[!htp]
    	\centering
   \includegraphics[width =\linewidth]{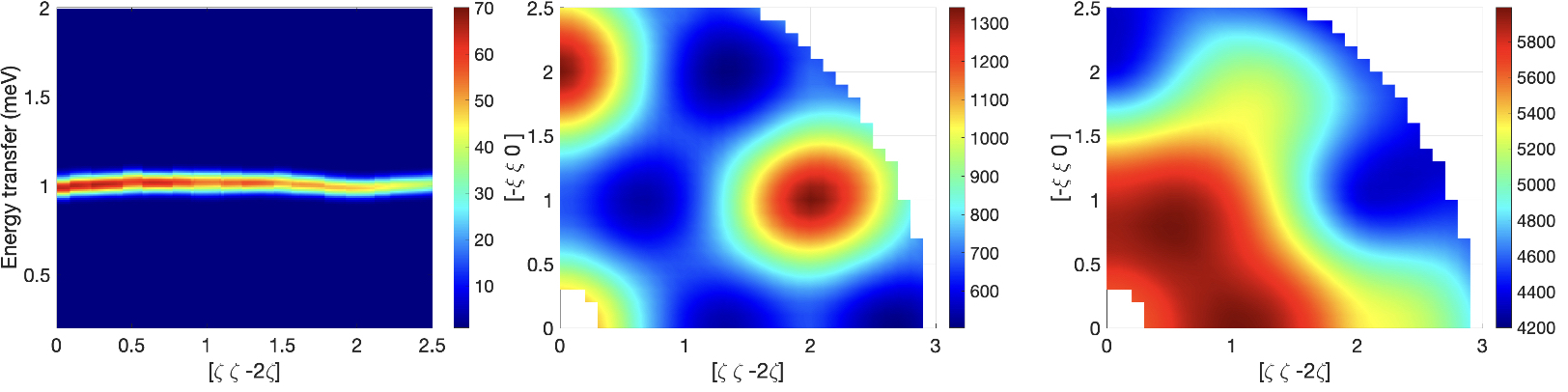}
		\caption{LSWT S(\textbf{Q}, $\omega$) and S(\textbf{Q}) profiles with $D$ = 110 mK and $J_{3}$ = -15 mK, (middle) low energy mode, 0.85 $\leq$ $\Delta$E $\leq$ 0.95 meV and (right) high energy mode 1.0 $\leq$ $\Delta$E $\leq$ 1.1 meV.}
        \label{fig:SQW_SQQ_DJ3only}
\end{figure} 

\begin{figure}[!htp]
    	\centering
   \includegraphics[width =\linewidth]{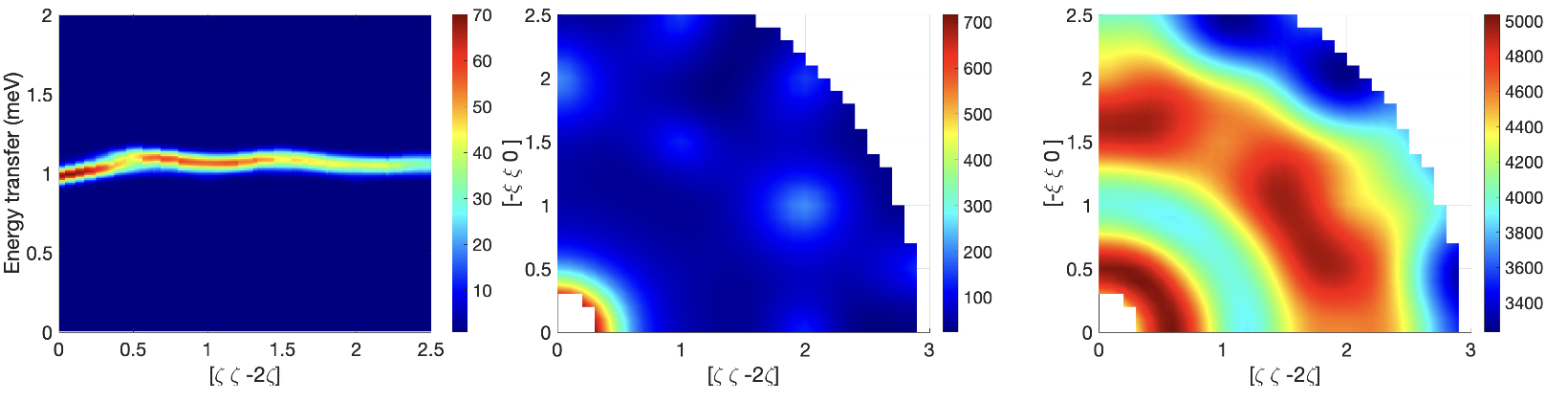}
		\caption{LSWT S(\textbf{Q}, $\omega$) and S(\textbf{Q}) profiles with $D$ = 110 mK, $J_{2}$ = -15 mK and $J_{3}$ = -15 mK, (middle) low energy mode, 0.85 $\leq$ $\Delta$E $\leq$ 0.95 meV and (right) high energy mode 1.0 $\leq$ $\Delta$E $\leq$ 1.1 meV.}
        \label{fig:SQW_SQQ_DJ2J3only}
\end{figure} 

\begin{figure}[!htp]
    	\centering
   \includegraphics[width =\linewidth]{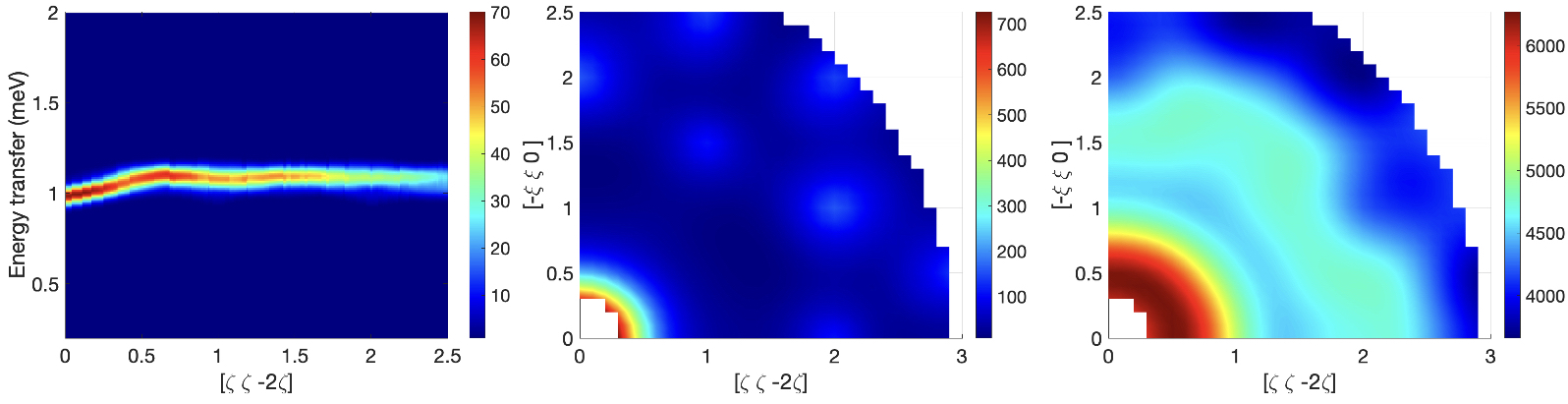}
		\caption{LSWT S(\textbf{Q}, $\omega$) and S(\textbf{Q}) profiles with $D$ = 110 mK, $J_{1}$ = -15 mK and $J_{2}$ = -15 mK, (middle) low energy mode, 0.85 $\leq$ $\Delta$E $\leq$ 0.95 meV and (right) high energy mode 1.0 $\leq$ $\Delta$E $\leq$ 1.1 meV.}
        \label{fig:SQW_SQQ_DJ1J2only}
\end{figure}

\begin{figure}[!htp]
    	\centering
   \includegraphics[width =\linewidth]{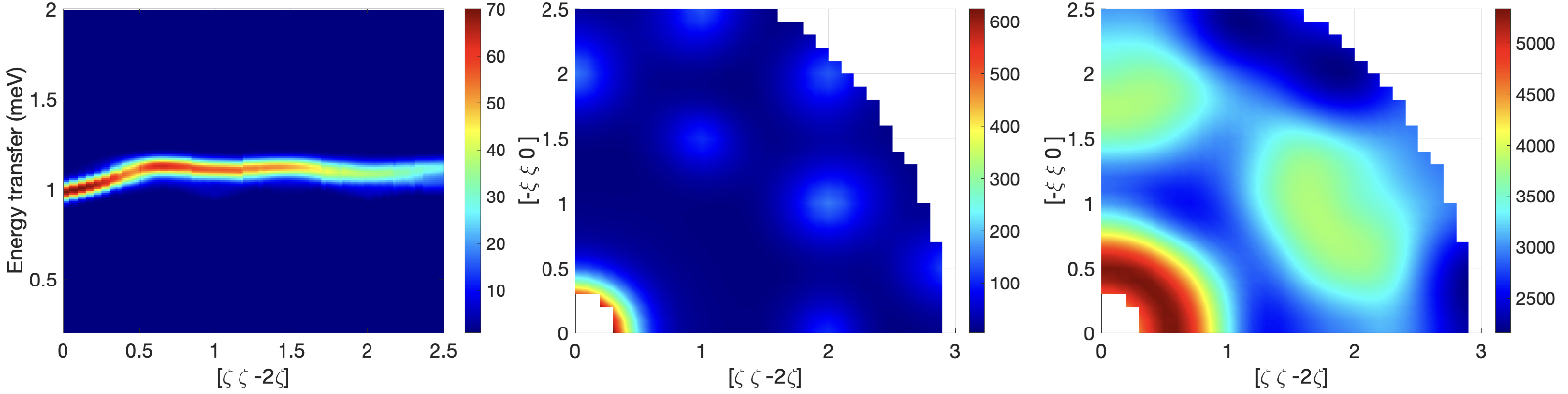}
		\caption{LSWT S(\textbf{Q}, $\omega$) and S(\textbf{Q}) profiles with $D$ = 110 mK, $J_{1}$ = -15 mK, $J_{2}$ = -15 mK and $J_{3}$ = -15 mK, (middle) low energy mode, 0.85 $\leq$ $\Delta$E $\leq$ 0.95 meV and (right) high energy mode 1.0 $\leq$ $\Delta$E $\leq$ 1.1 meV.}
        \label{fig:SQW_SQQ_DJ1J2J3only}
\end{figure}

\bibliography{YbGG_2023_FieldDep_2}

\end{document}